\documentclass[twocolumn]{webofc}

\usepackage[varg]{txfonts}   
\usepackage{amsmath} 
\usepackage{slashed}
\usepackage{braket}
\usepackage{bm}
\usepackage{graphicx} 
\usepackage{verbatim} 
\usepackage{color} 
\usepackage{subfigure} 
\usepackage{hyperref} 
\usepackage{longtable}
\usepackage{parskip} 
\usepackage[left]{lineno}
\usepackage{xcolor}

\usepackage[singlelinecheck=false 
]{caption}

\begin{document}
\title{Neutron star structure with nuclear force \\ mediated by hypothetical X17 boson} 

\author{\firstname{Vlasios} \lastname{Petousis}\inst{1}\fnsep\thanks{\email{vlasios.petousis@cern.ch}} \and
        \firstname{Martin} \lastname{Veselsk\'y}\inst{1} \and
        \firstname{Jozef} \lastname{Leja}\inst{2} 
}

\institute{Czech Technical University in Prague - Institute of Experimental and Applied Physics
\and
          Faculty of Mechanical Engineering - Slovak University of Technology in Bratislava
}

\abstract{
A reported ${17~}$MeV boson, which has been proposed as an explanation to the  $^{8}$Be and  $^{4}$He anomaly, is investigated in the context of its possible influence to neutron stars structure. Implementing a $m_{X}$=17 MeV to the nuclear equation of state using different incompressibility values K$_{0}$=245 MeV and K$_{0}$=260 MeV and solving Tolman-Oppenheimer-Volkoff equations, we estimate an upper limit of ${M_{TOV}\thickapprox 2.4M\odot}$ for a non rotating neutron star with span in radius ${R}$ between ${11.5~}$km to ${14~}$km. Moving away from pure - NN with admixture of 10\% protons and simulating possible softening of equation of state due to hyperons, we see that our estimated limits fit quite well inside the newest reported studies, coming from neutron stars merger event, GW190814
}

\maketitle

\section{Introduction}
In 2016 an article of Krasznahorkay et al. appeared \cite{Attila1}, where an anomaly in angular correlation of electron-positron decay of the 1$^{+}$ excited level of $^{8}$Be nucleus at ${18.15~}$MeV, specifically observed enhancement at folding angles close to 140 degrees, is interpreted as a signature of a decay via emission of neutral boson with mass around ${m_{X}=17~}$MeV. Since initial article \cite{Attila1}, similar effect was reported by the same group concerning a lower 1$^{+}$ excited state of $^{8}$Be at ${17.6~}$MeV \cite{Attila2} and more recently in the 0$^{-}$ excited state of $^{4}$He at ${21.01~}$MeV \cite{Attila4He}, reporting an enhancement at the folding angles close to 115 degrees. 
Through a recent work \cite{VPL}, we came up with an idea to investigate the effect of hypothetical ${17~}$MeV boson on nuclear matter and its influence on the structure of the compact astrophysical objects like neutron stars. The knowledge learned from the latest analysis on neutron stars, allows us to conclude that we know their mass within a good precision \cite{Lattimer,Cromartie}, but information on their radii is not so well defined yet \cite{Ozel, Guillot}. Encouragingly, recent observations of NICER \cite{Riley} neutron star radius, appeared with good accuracy and in future the mass-radius relation knowledge will be improved from the eXTP mission data \cite{Watts}.

In our investigation we examine the influence of equation of state assuming mediation of nuclear force by X17 boson on neutron star structure in case of pure neutron matter using two different incompressibility values K$_{0}$=245 MeV and K$_{0}$=260 MeV. Solving Tolman-Oppenheimer-Volkoff (TOV) equations for that equation of state we set up a limit in mass-radius relation for neutron stars. In our studies, we did not consider the case of a rotational neutron star which according to recent investigations done by others \cite{Suleimanov2020}, indicate possible overestimation of radii in non-rotating case by 3 - 3.5 km. 
First, we worked with pure neutron matter but also investigated possible admixture of protons. We investigated also possible softening of equation of state, an expected effect of hyperons, in order to attribute to the hyperons puzzle. In particular, our calculation indicates an upper limit of  ${M_{TOV}\thickapprox 2.4M\odot}$ neutron star with span in radius ${R}$ between  ${11.5~}$km to ${14~}$km. Concerning hyperons simulation, our findings shown a reduction in mass (down to ${M_{TOV}\thickapprox 2M\odot}$), as other relevant researches also indicate \cite{Fortin2018}. All our analysis has been performed for the aforementioned incompressibility values.

Our investigation indicates that equations of state used in present case modify significantly the internal structure of neutron stars. 
These resulting properties such as mass-radius relation fit quite well inside the newest reported studies as presented in papers \cite{Fattoyev2018, Fattoyev2020}, including also tidal deformability in neutron star radius coming from the observed neutron stars merger events. 
The paper is organized a s follows: in section 2 we introduce the equations of state for the assumption that nuclear force is being mediated by a ${17~}$MeV boson and model  of non-rotating neutron star using TOV equation, in section 3 we present our findings after the analysis  on the mass-radius and the density pressure correlations and in section 4 we discuss the results. 

\section{Tolman-Oppenheimer-Volkoff equations and the equation of state}
Going from the equations of states - ${\epsilon}$ and ${P}$ (the energy density and pressure at radius ${r}$) to a predicted mass - radius relation ${R(M)}$, requires integration of stellar structure equations. In general relativity, these are known as the TOV (Tolman-Oppenheimer-Volkoff) equations:

\begin{equation}
\frac{dP}{dr} = \frac{-G}{c^{2}}\frac{(P+\epsilon)(m+\frac{4 \pi r^{3}P}{c^{2}})}{r(r-\frac{2Gm}{c^{2}})} 
\end{equation}
\begin{equation}
\frac{dm}{dr} = 4\pi r^{2} \frac{\epsilon}{c^{2}} 
\end{equation}

where ${m(r)}$ is the total energy contained within radius ${r}$ and pressure ${P}$.
The energy density - ${\epsilon}$ and the pressure ${P}$ of pure neutron matter can be calculated using the formulae \cite{Serot}:

\begin{equation}
\begin{split}
\epsilon = \frac{g_{v}^2}{2 m_{v}^2} \rho_{N}^2 + \frac{g_{s}^2}{2 m_{s}^2} (m_{N}-m_{N}^{*})^2 + \frac{\kappa}{6 g_{s}^3} (m_{N}-m_{N}^{*})^3 \\
+ \frac{\lambda}{24 g_{s}^4} (m_{N}-m_{N}^{*})^4 + \frac{\gamma}{(2 \pi)^3} \int_{0}^{k_{F}} d^{3}k \sqrt{k^2 + (m_{N}^{*})^{2}}  \label{EQHDI}
\end{split}
\end{equation} 

\begin{equation}
\begin{split}
P = \frac{g_{v}^2}{2 m_{v}^2} \rho_{N}^2 + \frac{g_{s}^2}{2 m_{s}^2} (m_{N}-m_{N}^{*})^2 + \frac{\kappa}{6 g_{s}^3} (m_{N}-m_{N}^{*})^3 \\
+ \frac{\lambda}{24 g_{s}^4} (m_{N}-m_{N}^{*})^4 + \frac{1}{3} \frac{\gamma}{(2 \pi)^3} \int_{0}^{k_{F}} d^{3}k \frac{k^{2}} {\sqrt{k^{2} + (m_{N}^{*})^{2}}}  \label{EQHDI}
\end{split}
\end{equation}

where $g_{s}$,$m_{s}$ are coupling and rest mass of scalar boson,  $g_{v}$,$m_{v}$ are coupling and rest mass of vector boson, 
$\kappa$,$\lambda$ are couplings of cubic and quartic  self-interaction of scalar boson, $m_{N}$,$m_{N}^{*}$ are  rest mass of nucleon and its effective mass,  $\rho_{N}$ is the nucleonic density, $k_{F}$ is Fermi momentum of nucleons at zero temperature and  $\gamma$ is the degeneracy ($\gamma$=2 for neutron matter). The equations of state (Eq.3 and Eq.4) used here for TOV calculations are based on assumption that nuclear force  is being mediated by a ${17~}$MeV  boson possibly observed in the study of anomalous electron-positron pair production in excited states of $^{8}$Be \cite{Attila1,Attila2} and $^{4}$He \cite{Attila4He}.

In the simplest variant, without scalar boson self-interaction, the energy density is sensitive only on ratios $g_{s}/m_{s}$ and $g_{v}/m_{v}$. 
The corresponding couplings $g_{s}$ and $g_{v}$ are constrained using experimental binding energy values of symmetric nuclear matter ($E/A$=-16 MeV) and nuclear saturation  density ($\rho_{0}$=0.16$ fm^{-3}$), which are deduced from the  masses and radii of finite nuclei. The final constraints are  \cite{Serot}: 

\begin{equation}
\begin{split}
(\frac{m_{N}}{m_{s}})^2 g_{s}^2 = 357.4 \\
(\frac{m_{N}}{m_{v}})^2 g_{v}^2 = 273.8 \label{QHDIRatios}
\end{split}
\end{equation}

which guarantee proper values of both properties of nuclear matter.  However, such simple parameterization usually leads to unrealistically high values of incompressibility. 

In order to obtain more realistic values of incompressibility, it is necessary to implement  also cubic and quartic self-interaction of scalar meson. 
Using as a starting point the most promising values  of $\kappa$ from scan with cubic self-interaction term only between (${-0.0032}$ and ${0.0032~}$MeV), a range of $\lambda$ values was examined and ultimately regions of parameters were identified, allowing to obtain equations of state with values of incompressibility in range between K$_{0}$=240 and ${260~}$MeV (see Table \ref{tab1}). Similar values of imcompressibilty were extracted recently \cite{BinNSMerg} from analysis of  binary neutron star merger event GW170817 \cite{Abbott2017}. 
The resulting parameter sets for equation of states with incompressibilities K$_{0}$=245 MeV and K$_{0}$=260 MeV, respectively, are shown in Table \ref{tab1}. The mass of corresponding scalar particle does not correspond to any observed particle and it can be considered as an artefact of a given model. 

\begin{table}
\centering
\footnotesize
\caption{Constrained parameter sets for equations of state with incompressibilities K$_{0}$=245 MeV and K$_{0}$=260 MeV.}
\begin{tabular}{|l|c|c|c|c|c|c|}
\hline
K$_0$&m$_{v}$&m$_{s}$&g$_{v}$&g$_{s}$&$\kappa$&$\lambda$ \\
\hline
MeV & MeV& MeV& & & MeV& \\
\hline
245&17&25.58&0.2407&0.4666&0.0039&-0.001396 \\
260&17&25.58&0.2417&0.4684&0.00374&-0.001204 \\
\hline
\end{tabular}
\label{tab1}
\end{table}

In our recent work \cite{VPL} we further demonstrated that a hypothetical 
17 MeV boson fits some theory relations of chiral symmetry and 
instanton theory when dynamical quark mass is replaced by values 
close to quarks current masses, and corresponding couplings are close to 
these in Table \ref{tab1}. Referring to our recent work, we investigated that compared to
experimental pion mass around 
$m_{\pi}$=135 MeV the mass of pseudoscalar particle $m_{X}$=17 MeV 
means that, a ratio 
of dynamical quark mass around $m_{q,dyn}$=310 MeV to current quark 
mass of around $m_{q,curr}$=5 MeV fits the following relation almost perfectly:  
\begin{equation}
\frac{m_{X}^2} {m_{q,curr}} \simeq \frac{m_{\pi}^2} {m_{q,dyn}}
\label{MassScaling}
\end{equation}

This appears to signal that as explicit breaking of chiral symmetry 
is restored from dynamical mass scale down to current mass scale, 
and also the properties of pseudo-scalar particle get 
closer to Goldstone boson. 
The relation (\ref{MassScaling}) then suggests that same 
outcome is reached when considering sum of axial current of current quark 
and axial current of 17 MeV boson. The 17 MeV boson thus appears 
to restore the symmetry of QCD in vacuum, where current 
quarks are considered as a degree of freedom.  
Based on this, it is possible to derive an analogue 
of the Goldberger-Treiman relation \cite{G-T}
\begin{equation}
g_{Xqq} = \frac{g_{A} m_{q,curr}}{f_{\pi}}
\label{GTXqq}
\end{equation}

where $g_{Xqq}$ is the coupling of $X$-boson to current quark, 
$g_{A}$ is renormalization factor of axial current and 
$f_{\pi}$ was adopted as proportionality factor also for 
axial current of $X$-boson. The value of coupling $g_{Xqq}$=0.07 is obtained, 
what suggests coupling with nucleon $g_{XNN}$=0.21
In this respect, it is apparent 
that also results using relativistic mean field theory 
of nuclear force, presented above, point practically in same 
direction, since the value of coupling corresponding 
to vector mesonic mass of 17 MeV is $g_v$=0.24.

This situation is depicted in the Fig.1, which compares two trends of mesonic versus quark masses. First trend combines formulas (\ref{QHDIRatios}) and (\ref{GTXqq}) under assumption that the obtained coupling correspond to ${{g_{XNN}} = {3g_{Xqq}}}$. Second trend represents formula (\ref{MassScaling}). Both trends cross around the value of $m_{X}$=17 MeV and $m_{q}$=5 MeV. 

Possible observation of a boson with mass $m_{X}$=17 MeV and 
relations (\ref{MassScaling}) and (\ref{GTXqq}) thus suggests the
existence of a new scale in QCD, with chiral symmetry 
practically restored. 
\begin{figure}
\centering
\includegraphics[width=87mm, height=65mm]{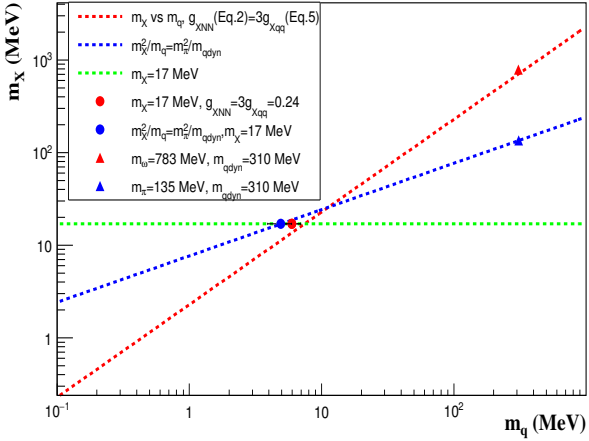}
\caption{(Color online). Trends of mesonic versus quark masses. First trend combines formulas (2) and (5) as explained in the text. Second trend represents formula (4).}
\end{figure}
The  reason for apparent restoration of chiral symmetry 
suggested above is a matter of discussion. It might be related to 
interaction between nucleons at distance, mediated by instantons, 
where possibly the density of instantons mediating interaction 
in a region between nucleons becomes smaller and  
possibly a bounce into a false instanton vacuum (a quantum mechanical tunneling into a metastable vacuum and back), 
a mechanism based on instanton model \cite{Coleman}, can occur. 

\section{Analysis results}
For Tolman-Oppenheimer-Volkoff calculations \cite{Joonas} the equations
of state ${P(\rho})$ needed to be expressed in form of polytropes.  Three transition densities are defined, $\rho_{1}$=2.8$\times$10$^{14}$ g/cm$^{3}$, $\rho_{2}$=10$^{14.7}$ g/cm$^{3}$, and $\rho_{3}$=10$^{15}$ g/cm$^{3}$,  and four parameters are calculated, three exponents of the power law polytropes  $\Gamma_{1},\Gamma_{2},\Gamma_{3}$, respectively,  and a value $a_{0}= log(p(\rho_{1}))+\Gamma_{1} (log(\rho_{2})-log(\rho_{1}))$. The resulting values are shown in Table \ref{tab2}. The pressure as function of nuclear density for all model parameters shown in Fig.2. 

\begin{table}
\centering
\footnotesize
\caption{Parameters of polytropes used as equations of state for the Tolman-Oppenheimer-Volkoff calculations.}
\begin{tabular}{|l|c|c|c|c|c|c|}
\hline
Models & $a_{0}$& $\Gamma_{1}$& $\Gamma_{2}$& $\Gamma_{3}$& K$_0$(MeV) \\
\hline
    PNM1  & 34.600 & 3.350 & 3.170 & 2.497 &245 \\
    H1 (Hyper) & 34.600 & 3.350 & 2.570 & 1.497 &245  \\
    p1 (10\%)  & 34.553 & 3.470 & 3.307 & 2.546 &245  \\  
  \hline
    PNM2  & 34.642 & 3.348 & 3.040 & 2.534 &260  \\
    H2 (Hyper) & 34.642 & 3.348 & 2.440 & 1.534 &260  \\
    p2 (10\%) & 34.593 & 3.530 & 3.177 & 2.500&260  \\
\hline
\end{tabular}
\label{tab2}
\end{table}

\begin{figure}
\centering
\includegraphics[width=80mm, height=65mm]{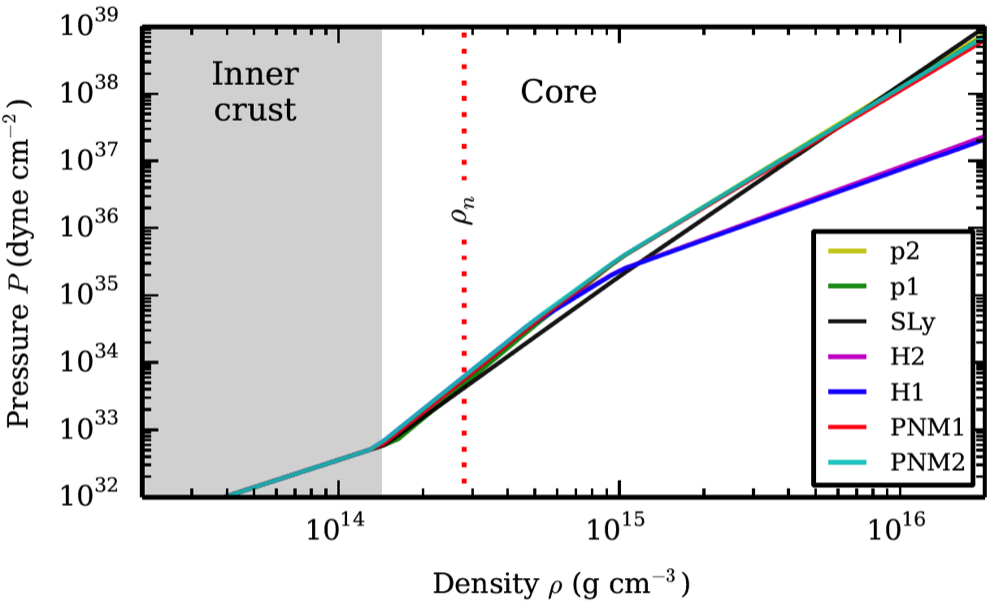}
\caption{(Color online). The pressure as function of nuclear density for all the model parameters as they defined in the Table \ref{tab2}, based on assumption that nuclear force  is being mediated by a ${17~}$MeV boson.}
\end{figure}

The polytropes named as PNM1 and PNM2 (Fig.3) correspond to equations  of state shown in Table \ref{tab2}, with incompressibilities K$_{0}$=245 MeV and K$_{0}$=260 MeV respectively. These equations of state represent properties of pure neutron matter. The maximum mass of neutron star reaches up to 2.4 of Solar masses, what is close to values 2.5 - 2.67 derived for the mass of secondary in the observed astrophysical event GW190814 \cite{GW190814}.  
However, the properties of neutron stars can be influenced by admixture of protons. During formation of neutron stars most of the protons are transformed into neutrons by inverse ${\beta}$-process, contributing significantly to cooling of neutron star. Still, some amount of  protons and electrons may remain present. 
The concentration of protons may vary in different layers of neutron star, depending on the local ${\beta}$ - equilibrium, most probably it can be expected significant at outer regions with lower densities, which apparently determine the radius of neutron star. 

We investigate sensitivity to proton admixture by assuming a fixed proton concentration of 10\% in the whole volume of neutron star. Coulomb interaction is assumed to be counterbalanced by equal concentration of electrons. The resulting polytropes are named as p1 and p2 (Fig.3) and the admixture of protons appears to modify the radius of neutron star moderately while the maximum mass remains constant.  The admixture of protons does not contradict possible observation of neutron star with mass of 2.6 times the Solar mass \cite{GW190814}. The radius of neutron star is reduced by a 10 \% admixture of protons by approximately 3 \%, which is relatively moderate change. Thus the properties of neutron star appear only moderately sensitive to the local ${\beta}$ - equilibrium. 

\begin{figure}
\centering
\includegraphics[width= 80mm, height=65mm]{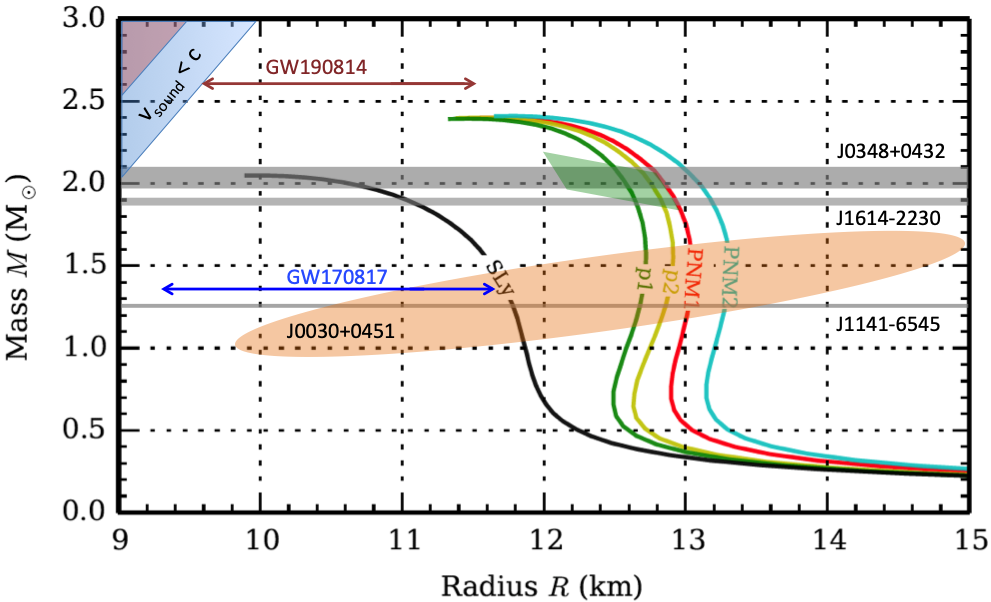}
\caption{(Color online). The mass-radius relation for pure neutron matter (PNM1 and PNM2) and with an admixture of 10\% protons (p1 and p2) for K$_{0}$=245 and ${260~}$MeV respectively. The calculations based on assumption that nuclear force is being mediated by a ${17~}$MeV boson.} 
\end{figure}

\begin{figure}
\centering
\includegraphics[width=80mm, height=65mm]{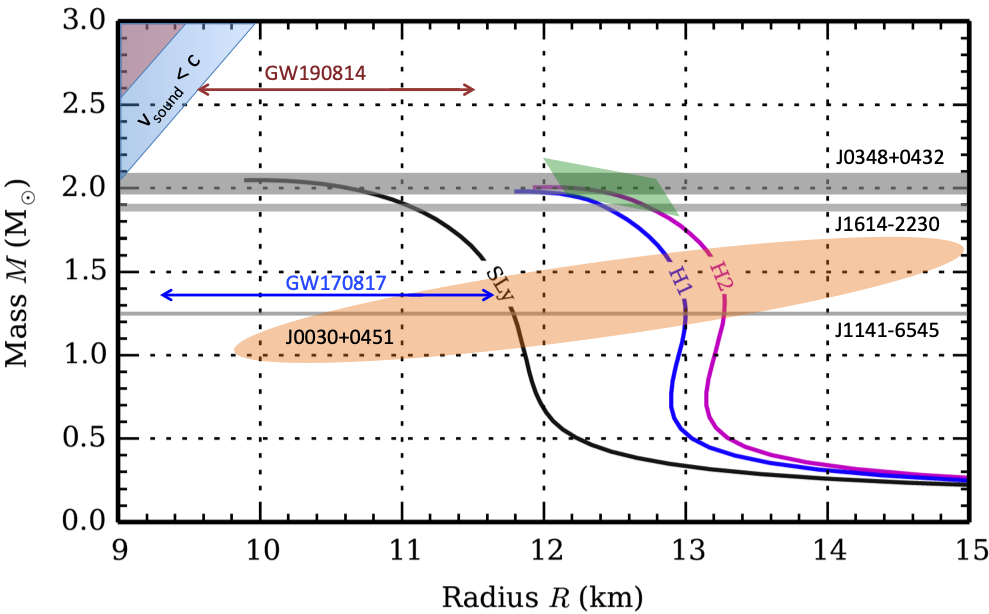}
\caption{(Color online). The mass-radius relation for neutron matter including hyperons (H1 and H2) for K$_{0}$=245 and ${260~}$MeV respectively. The green area shows the uncertainty on the mass and radius of stars with hyperons in the core due to the fact that the properties of these particles are barely constrained by laboratory measurements. The calculations based on assumption that nuclear force is being mediated by a ${17~}$MeV boson.}
\end{figure}

Since it can be expected that at densities above two times saturation density, the non-nucleonic particles such as  hyperons can play increasing role and soften the equation of state at these densities, such effect was simulated  by polytropes under acronyms H1 and H2. The equation of state is softened significantly at high densities while  the properties close to saturation density remain unchanged. Even such rough estimate may indicate the realistic behavior. 
The simulated effect of hyperons, represented by softening equation of state above two times the saturation density, translates mainly into decrease of maximum mass of neutron star, while the radius remains practically constant up to neutron star mass of 1.5 of Solar mass. 
A maximum mass around two times the Solar mass corresponds to maximum observed pulsar mass which was valid until very recently and thus the parameterizations H1 and H2 (Fig.4), represent maximum extent of such effect and no further softening of equation of state would be plausible. On the other hand, recent possible observation of neutron star with mass of 2.6 times the Solar mass \cite{GW190814} would apparently exclude any significant effect of hyperons at all. 

\section{Conclusions}
In summary, the properties of a non rotating neutron stars are simulated using the equation of state, based on assumption that nuclear force 
is being mediated by a 17 MeV boson possibly observed in the study of anomalous electron-positron pair production in excited states of $^{8}$Be \cite{Attila1,Attila2} and $^{4}$He \cite{Attila4He}. 
The presented analysis was performed using pure neutron matter and also an admixture of protons. We investigated also hyperons simulation in order to address the hyperons puzzle. Our investigation indicates an upper limit of  ${M_{TOV}\thickapprox 2.4M\odot}$ neutron stars with span in radius ${R}$ between ${11.5~}$km to ${14~}$km for normal neutron matter and moderate reduction of radii for the admixture with 10\% protons. 
Concerning the possible effect of hyperons, our findings shown reduction in maximum mass down to ${M_{TOV}\thickapprox 2M\odot}$. We conclude that our results appear in agreement with observed properties of neutron stars. In particular, the possible existence of neutron star with mass up to 2.6 times the Solar mass, reported recently \cite{GW190814}, appears plausible.  

\section{Ackwnoledgments}
This work is supported by the European Regional Development Fund-Project
Engineering applications of microworld physics, (Contract No.
CZ.02.1.01/0.0/0.0/16\_019/0000766).

\end{document}